\documentclass[aps,reprint,nofootinbib]{revtex4-1}

\usepackage[english]{babel}
\usepackage[utf8x]{inputenc}
\usepackage[T1]{fontenc}

\usepackage{float}
\usepackage{amsmath}
\usepackage{hyperref}
\usepackage{cleveref}
\usepackage{graphicx}
\usepackage{glossaries}

\usepackage{physics}
\usepackage{siunitx}

\newcommand{\delay}[1]{\mathbf{D}_{#1}}
\newcommand{\ddelay}[1]{\dot{\mathbf{D}}_{#1}}

\newcommand{\poly}[1]{\mathbf{P}_{#1}}

\newcommand{\fourier}[1]{\opbraces{\widetilde{#1}}}
\newcommand{\fourierddelay}[1]{\fourier{\dot{\mathbf{D}}}_{#1}}
\newcommand{\fourierpoly}[1]{\fourier{\mathbf{P}}_{#1}}

\newcommand{\psd}[1]{\opbraces{S_{#1}}}

\newcommand{\idxset}{\mathcal{I}}

\renewcommand{\qc}{\,\text{,}}
\newcommand{\qs}{\,\text{.}}

\newacronym{lisa}{LISA}{Laser Interferometer Space Antenna}
\newacronym{gw}{GW}{gravitational wave}
\newacronym{gpu}{GPU}{graphics processing unit}
\newacronym{esa}{ESA}{European Space Agency}
\newacronym{tdi}{TDI}{time-delay interferometry}
\newacronym{fir}{FIR}{finite impulse response}
\newacronym{iir}{IIR}{infinite impulse response}
\newacronym[longplural={power spectral densities}]{psd}{PSD}{power spectral density}
\newacronym[longplural={amplitude spectral densities}]{asd}{ASD}{amplitude spectral density}
\newacronym{ldc}{LDC}{LISA Data Challenge}
\newacronym{mldc}{MLDC}{Mock LISA Data Challenge}
\newacronym{ssb}{SSB}{Solar system's barycenter}
\newacronym{uflis}{UFLIS}{University of Florida LISA simulator}
\newacronym{lot}{LOT}{LISA On Table}
\newacronym{apc}{APC}{Astroparticules et Cosmologie}
\newacronym[longplural={movable optical system assemblies}]{mosa}{MOSA}{movable optical system assembly}
\newacronym{uso}{USO}{ultra-stable oscillator}
\newacronym{adc}{ADC}{analog-to-digital converter}
\newacronym{eom}{EOM}{electro-optical modulator}

\newacronym{isc}{ISC}{inter-spacecraft}
\newacronym{tm}{TM}{test-mass}
\newacronym{ref}{REF}{reference}

\begin{document}

\title{Clock-jitter reduction in LISA time-delay interferometry combinations}

\author{Olaf Hartwig}
\affiliation{Max-Planck-Institut für Gravitationsphysik (Albert-Einstein-Institut),
Callinstraße 38, 30167 Hannover, Germany}

\author{Jean-Baptiste Bayle}
\affiliation{Jet Propulsion Laboratory, California Institute of Technology, 4800 Oak Grove Drive, Pasadena, CA 91109, USA}

\date{\today}


\begin{abstract}
The Laser Interferometer Space Antenna (LISA) is a European Space Agency mission that aims to measure gravitational waves in the millihertz range. The three-spacecraft constellation forms a nearly-equilateral triangle, which experiences flexing along its orbit around the Sun. These time-varying and unequal armlengths require to process measurements with time-delay interferometry (TDI) to synthesize virtual equal-arm interferometers, and reduce the otherwise overwhelming laser frequency noise. Algorithms compatible with such TDI combinations have recently been proposed in order to suppress the phase fluctuations of the onboard ultra-stable oscillators (USO) used as reference clocks.

In this paper, we propose a new method to cancel USO noise in TDI combinations. This method has comparable performance to existing algorithms, but is more general as it can be applied to most TDI combinations found in the literature. We compute analytical expressions for the residual clock noise before and after correction, accounting for the effect of time-varying beatnote frequencies, previously neglected. We present results of numerical simulations that are in agreement with our models, and show that clock noise can be suppressed below required levels. The suppression algorithm introduces a new modulation noise, for which we propose a partial mitigation. This modulation noise remains the limiting effect for clock-noise suppression, setting strict timing requirements on the sideband generation.
\end{abstract}
\maketitle

\section{Introduction}

The \gls{lisa} is a \gls{esa} scientific space mission, which aims to measure \glspl{gw} in the millihertz range~\cite{Audley:2017drz}. Those waves are predicted by Einstein's theory of general relativity and produced by the quadrupolar moment of very dense objects, such as black hole binaries or coalescing super-massive black holes. The detection of low-frequency \glspl{gw} will help answer numerous astrophysical, cosmological, and theoretical questions.

The mission is expected to be launched in 2034. Three spacecraft will trail the Earth around the Sun, in a nearly equilateral triangular configuration with armlengths of about 2.5 million kilometers. Each spacecraft contains two free-falling test masses acting as inertial sensors, with a demonstrated sub-femto-$g$ accuracy~\cite{Armano:2016bkm,Armano:2018kix}. The relative motions of these test masses is measured to picometer precision using laser interferometry. These measurements are then processed on ground to extract \gls{gw} signals. Each spacecraft carries one \gls{uso}, which provides a unique clock time, used as reference for all systems of the onboard measurement chain. Any timing jitter of the onboard clocks will couple into our interferometric measurements. State-of-the-art space-qualified oscillators typically have an Allan deviation of about \num{e-13} for averaging times above \SI{1}{\second}~\cite{Weaver2010}. This is not sufficient to achieve the picometer test-mass position readout required by \gls{lisa}.

As a solution, it is planned to perform an independent measurement of the differential clock jitter~\cite{Barke:2015}, which can be used for clock-noise reduction algorithms. A first version of such an algorithm was presented in~\cite{Hellings:2001}, which perfectly cancels clock noise assuming constant armlengths. In~\cite{Tinto:2018}, it was shown that this algorithm can be extended to linearly time-varying armlengths and still reduces clock noise below requirements. This correction algorithm is designed to be applied to the so-called \gls{tdi} combinations, proposed in ~\cite{Tinto:1999yr} to reduce the otherwise overwhelming laser frequency noise. \Gls{tdi} combinations synthesize virtual laser noise-free interferometers of different topologies. Under the assumption of constant armlengths, it has been shown~\cite{Dhurandhar:2002zcl} that all topologies can be generated from a finite set of combinations. However, analytic and numerical studies~\cite{Vallisneri:2004bn,Petiteau:2008zz,Bayle:2019,Muratore:2020mdf} have shown that these first-generation combinations are not enough to reach the required laser-noise reduction. Instead, one must use second-generation combinations, for which no finite set of generators has been found~\cite{tinto2020}. Each of these combinations might be useful for scientific analysis as it possesses a unique response to instrumental noise and gravitational signals. The previously mentioned correction algorithm is only applicable to a specific subset of solutions, namely the Michelson and Sagnac topologies. We present here a general clock-noise reduction algorithm which can be applied to almost any second-generation \gls{tdi} combination.

In \cref{sec:phasemeter-measurements}, we introduce a model for the phasemeter measurements produced by \gls{lisa}. Then, in \cref{sec:clock-noise-reduction}, we study how clock noise enters in the standard \gls{tdi} combinations, and propose a general algorithm to remove it. In \cref{sec:simulations}, we present numerical simulations, and discuss the main results in \cref{sec:discussion}. In particular, we give models for the limiting effects and compare the performance of our algorithm against existing clock-noise reduction schemes. We conclude in \cref{sec:conclusion}.

\section{Phasemeter measurements}
\label{sec:phasemeter-measurements}

\subsection{Description of laser beams}
\label{ssec:description-of-laser-beams}

\begin{figure}
    \centering
    \includegraphics[width=\columnwidth]{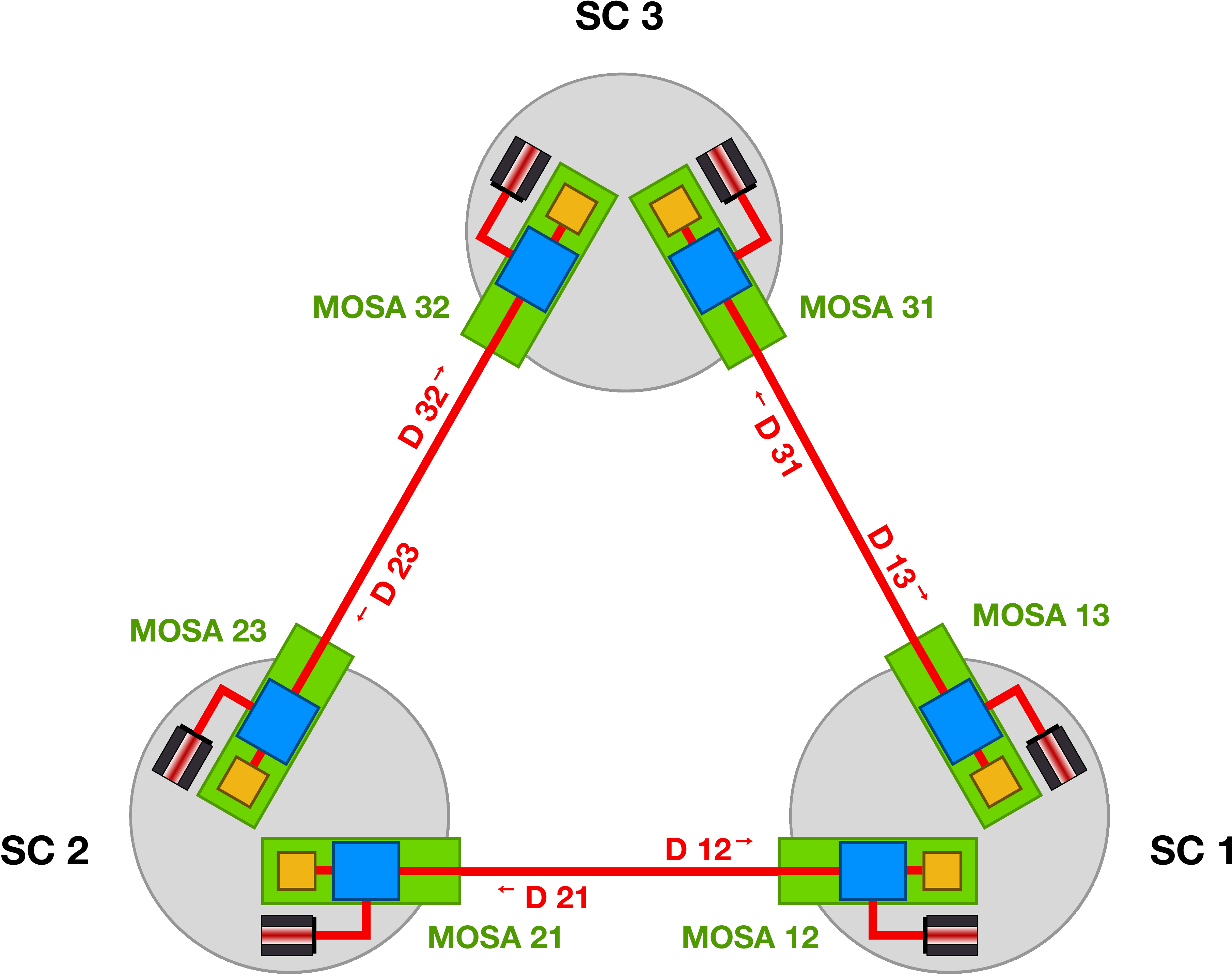}
    \caption{Labelling conventions used for spacecraft, light travel times, lasers, \glspl{mosa}, and interferometric measurements.}
    \label{fig:labelling}
\end{figure}

We give here a description of our instrumental model, using the latest recommendations on conventions and notations established by the \gls{lisa} Consortium (illustrated in \cref{fig:labelling}).

Spacecraft are indexed 1, 2, 3 clockwise when looking down at their solar panels. Any subsystem or measurement uniquely attached to a spacecraft is labelled with the same index. E.g., each spacecraft hosts a single clock.

\Glspl{mosa} are labelled with two indices $ij$. The former matches the index~$i$ of the spacecraft hosting the \gls{mosa}, while the second index is that of the spacecraft~$j$ exchanging light with the \gls{mosa}. Any subsystem or measurement uniquely attached to a \gls{mosa} shares the same indices.

In the following, we make use of two sets of time coordinates. On one hand, the spacecraft proper time $\tau_i$ is used to describe all physical signals in spacecraft $i$. On the other hand, each spacecraft hosts a clock implemented by an \gls{uso}. Instrumental imperfections of this \gls{uso} are modeled by the clock timing jitter $q_i(\tau)$, measured with respect to the spacecraft proper time. This defines the clock time coordinate
\begin{equation}
    \hat\tau_i(\tau) = \tau + q_i(\tau)
    \qs
    \label{eq:clock-time-definition}
\end{equation}
We neglect any deterministic clock imperfections, such as a constant frequency offset, and assume that $q_i$ can be modelled as zero-mean Gaussian noise with a \gls{psd} given in \cref{eq:clock-noise-psd}.

In order to reduce clock noise, we transfer clock tones between the spacecraft. Clock information is imprinted on the laser beams via a phase modulation at \si{\giga\hertz} frequencies. Because the modulation index is small (around \num{0.15}), we can describe the modulated laser beam as the superposition of three independent beams, each with its own total frequency: the carrier frequency $\nu_\text{c}(\tau)$, the upper sideband frequency, and the lower sideband frequency. In the following, we assume that the information contained in both sidebands is identical, and will only model the upper sideband, designated simply as \textit{the sideband}. Its total frequency is denoted $\nu_\text{sb}(\tau)$.

We decompose the carrier and sideband total frequencies into a constant nominal laser frequency $\nu_0 = \SI{281.6}{\tera\hertz}$, large out-of-band offsets $\nu^o(\tau)$, and small in-band fluctuations $\nu^\epsilon(\tau)$,
\begin{subequations}
\begin{align}
    \nu_\text{c}(\tau) &= \nu_0 + \nu_\text{c}^o(\tau) + \nu_\text{c}^\epsilon(\tau)
    \qc
    \label{eq:carrier-decomposition}
    \\
    \nu_\text{sb}(\tau) &= \nu_0 + \nu_\text{sb}^o(\tau) + \nu_\text{sb}^\epsilon(\tau)
    \qs
    \label{eq:sideband-decomposition}
\end{align}
\end{subequations}
Frequency offsets contain Doppler shifts~\cite{Bayle:2021mue}, the tens-of-\si{\mega\hertz} programmed offsets as determined by the frequency plan, and constant sideband frequency offsets $\nu_{ij}^m \approx \SI{2.4}{\giga\hertz}$\footnote{In this paper, we assume that on each spacecraft, the left-handed modulation signals are at exactly \SI{2.4}{\giga\hertz}, while the right-handed ones are at \SI{2.401}{\giga\hertz}.}. Frequency fluctuations are used to model instrumental noise.

In this paper, we assume all six lasers to be independent. On~\gls{mosa} $ij$, the laser source~$ij$ delivers the \textit{local carrier and sideband} with respective frequencies of $\nu_{ij,\text{c}}(\tau)$ and $\nu_{ij,\text{sb}}(\tau)$, which we decompose into offsets and fluctuations according to \cref{eq:carrier-decomposition,eq:sideband-decomposition},
\begin{subequations}
\begin{align}
    \nu_{ij,\text{c}}^o(\tau) &= O_{ij}
    \qc
    \label{eq:local-carrier-offsets}
    \\
    \nu_{ij,\text{c}}^\epsilon(\tau) &= 0
    \qc
    \label{eq:local-carrier-fluctuations}
    \\
    \nu_{ij,\text{sb}}^o(\tau) &= O_{ij} + \nu_{ij}^m
    \qc
    \label{eq:local-sideband-offsets}
    \\
    \nu_{ij,\text{sb}}^\epsilon(\tau) &= \nu_{ij}^m (\dot q_i(\tau) + \dot N^m_{ij}(\tau))
    \qs
    \label{eq:local-sideband-fluctuations}
\end{align}
\end{subequations}
We assume here that the programmed frequency offsets $O_{ij}$ are constant. In addition, we assume that the traditional \gls{tdi} algorithms sufficiently suppress primary noises (e.g., laser noise or spacecraft jitter) such that we can neglect any noises apart from the clock timing jitter. In addition, we account for any imperfections in the sideband modulation chain, such as errors due to electrical-frequency conversions or optical modulators, by a modulation noise term $N^m_{ij}(\tau)$. Both $N^m_{ij}(\tau)$ and $\dot q_i(\tau)$ are expressed as fractional frequency fluctuations, scaled by the modulation frequency $\nu_{ij}^m$.

On \gls{mosa}~$ij$, we call \textit{distant carrier and sideband} the beams received from the distant \gls{mosa}~$ji$. We define the delay operator associated with the light travel time\footnote{In all generality, delay operators must include the conversion between the emitting and receiving spacecraft proper time coordinates. We include these corrections based on a derivation provided by the SYRTE “Theory and Metrology” group.} $d_{ij}(\tau)$ between spacecraft $j$ and $i$ as 
\begin{equation}
    \delay{ij} x(\tau) = x(t - d_{ij}(\tau))
    \qc
\end{equation}
for any signal $x(\tau)$. Because these light travel times are time-varying, we have to account for Doppler shifts proportional to the time derivatives of the light travel times $\dot d_{ij}(\tau)$~\cite{Bayle:2021mue}. Therefore, we define the Doppler-delay operator,
\begin{equation}
    \ddelay{ij} x(\tau) = (1 - \dot d_{ij}(\tau)) x(t - d_{ij}(\tau))
    \qs
\end{equation}

Onboard \gls{mosa}~$ij$, the distant carrier and sideband are then given by
\begin{subequations}
\begin{align}
    \nu_{ij \leftarrow ji,\text{c}}(\tau) &= \ddelay{ij} \nu_{ji,\text{c}}(\tau)
    \qc
    \\
    \nu_{ij \leftarrow ji,\text{sb}}(\tau) &= \ddelay{ij} \nu_{ji,\text{sb}}(\tau)
    \qs
\end{align}
\end{subequations}

We decompose again these frequencies into offsets and fluctuations. Using \cref{eq:carrier-decomposition,eq:sideband-decomposition,eq:local-carrier-offsets,eq:local-carrier-fluctuations,eq:local-sideband-offsets,eq:local-sideband-fluctuations}, we have
\begin{subequations}
\begin{align}
    \nu_{ij \leftarrow ji,\text{c}}^o(\tau) &= \ddelay{ij} O_{ji} - \nu_0 \dot d_{ij}(\tau)
    \qc
    \label{eq:distant-carrier-offsets}
    \\
    \nu_{ij \leftarrow ji,\text{c}}^\epsilon(\tau) &= 0
    \qc
    \label{eq:distant-carrier-fluctuations}
    \\
    \nu_{ij \leftarrow ji,\text{sb}}^o(\tau) &= \ddelay{ij} (O_{ji} + \nu^m_{ji}) - \nu_0 \dot d_{ij}(\tau)
    \qc
    \label{eq:distant-sideband-offsets}
    \\
    \nu_{ij \leftarrow ji,\text{sb}}^\epsilon(\tau) &= \nu^m_{ji} \ddelay{ij} (\dot q_j(\tau) + \dot N^m_{ji}(\tau))
    \qs
    \label{eq:distant-sideband-fluctuations}
\end{align}
\end{subequations}

Local laser beams are also exchanged with adjacent \glspl{mosa}. On \gls{mosa}~$ij$, we call the beams received from adjacent \gls{mosa}~$ik$ the \textit{adjacent carrier and sideband}. We ignore any effect due to the propagation of the beams inside a spacecraft, such that their frequencies decomposed into offsets and fluctuations are simply given as
\begin{subequations}
\begin{align}
    \nu_{ij \leftarrow ik,\text{c}}^o(\tau) &= O_{ik}
    \qc
    \label{eq:adjacent-carrier-offsets}
    \\
    \nu_{ij \leftarrow ik,\text{c}}^\epsilon(\tau) &= 0
    \qc
    \label{eq:adjacent-carrier-fluctuations}
    \\
    \nu_{ij \leftarrow ik,\text{sb}}^o(\tau) &= O_{ik} + \nu^m_{ik}
    \qc
    \label{eq:adjacent-sideband-offsets}
    \\
    \nu_{ij \leftarrow ik,\text{sb}}^\epsilon(\tau) &= \nu^m_{ik}  (\dot q_i(\tau) + \dot N^m_{ik}(\tau))
    \qs
    \label{eq:adjacent-sideband-fluctuations}
\end{align}
\end{subequations}

\subsection{Interferometric measurements}
\label{ssec:interferometric-measurements}

Onboard each \gls{mosa}, three heterodyne interferometers measure beatnotes frequencies $\nu_\text{BN}(\tau)$. The photodetectors recording these beatnotes have a limited bandwidth between \SI{5}{\mega\hertz} and \SI{25}{\mega\hertz}, such that only three beatnotes are detected in each interferometer: one between the two carriers of the incoming beams, one between the upper sidebands and one between the lower sidebands. We give here only expressions for the carrier-carrier $\nu_{\text{BN},\text{c}}(\tau)$ and upper sideband-upper sideband $\nu_{\text{BN},\text{sb}}(\tau)$ beatnotes.

Following the formalism introduced in \cref{eq:carrier-decomposition,eq:sideband-decomposition}, we decompose the beatnote frequencies into large out-of-band offsets and small in-band fluctuations, $\nu_\text{BN}(\tau) = \nu_\text{BN}^o(\tau) + \nu_\text{BN}^\epsilon(\tau)$. If we denote the two incoming beams 1 and 2, we have
\begin{subequations}
\begin{align}
    \nu_{\text{BN},\text{c}}^o(\tau) &= \nu_{1,\text{c}}^o(\tau) - \nu_{2,\text{c}}^o(\tau)
    \qc
    \\
    \nu_{\text{BN},\text{c}}^\epsilon(\tau) &= \nu_{1,\text{c}}^\epsilon(\tau) - \nu_{2,\text{c}}^\epsilon(\tau)
    \qc
    \\
    \nu_{\text{BN},\text{sb}}^o(\tau) &= \nu_{1,\text{sb}}^o(\tau) - \nu_{2,\text{sb}}^o(\tau)
    \qc
    \\
    \nu_{\text{BN},\text{sb}}^\epsilon(\tau) &= \nu_{1,\text{sb}}^\epsilon(\tau) - \nu_{2,\text{sb}}^\epsilon(\tau)
    \qs
\end{align}
\end{subequations}

The \textit{\gls{isc} interferometer} mixes the local and distant beams. Using \cref{eq:local-carrier-offsets,eq:local-carrier-fluctuations,eq:local-sideband-offsets,eq:local-sideband-fluctuations} and \cref{eq:distant-carrier-offsets,eq:distant-carrier-fluctuations,eq:distant-sideband-offsets,eq:distant-sideband-fluctuations}, we find
\begin{subequations}
\begin{align}
    \nu_{\text{isc}_{ij},\text{c}}^o(\tau) &= \ddelay{ij} O_{ji} - O_{ij} - \nu_0 \dot d_{ij}(\tau)
    \qc
    \label{eq:isc-carrier-offsets}
    \\
    \nu_{\text{isc}_{ij},\text{c}}^\epsilon(\tau) &= 0
    \qc
    \label{eq:isc-carrier-fluctuations}
    \\
    \nu_{\text{isc}_{ij},\text{sb}}^o(\tau) &=  (\ddelay{ij} O_{ji} - O_{ij} - \nu_0 \dot d_{ij}(\tau)) + (\ddelay{ij} \nu^m_{ji} - \nu^m_{ij})
    \qc
    \label{eq:isc-sideband-offsets}
    \\
    \nu_{\text{isc}_{ij},\text{sb}}^\epsilon(\tau) &= \nu^m_{ji} \ddelay{ij} (\dot q_j(\tau) + \dot N^m_{ji}(\tau)) - \nu_{ij}^m (\dot q_i(\tau) + \dot N^m_{ij}(\tau))
    \qs
    \label{eq:isc-sideband-fluctuations}
\end{align}
\end{subequations}

The \textit{\gls{ref} interferometer} mixes the local and adjacent beams. Using \cref{eq:local-carrier-offsets,eq:local-carrier-fluctuations,eq:local-sideband-offsets,eq:local-sideband-fluctuations} and \cref{eq:adjacent-carrier-offsets,eq:adjacent-carrier-fluctuations,eq:adjacent-sideband-offsets,eq:adjacent-sideband-fluctuations},
\begin{subequations}
\begin{align}
    \nu_{\text{ref}_{ij},\text{c}}^o(\tau) &= O_{ik} - O_{ij}
    \qc
    \label{eq:ref-carrier-offsets}
    \\
    \nu_{\text{ref}_{ij},\text{c}}^\epsilon(\tau) &= 0
    \qc
    \label{eq:ref-carrier-fluctuations}
    \\
    \nu_{\text{ref}_{ij},\text{sb}}^o(\tau) &= (O_{ik} - O_{ij}) + (\nu^m_{ik} - \nu^m_{ij})
    \qc
    \label{eq:ref-sideband-offsets}
    \\
    \nu_{\text{ref}_{ij},\text{sb}}^\epsilon(\tau) &= \nu^m_{ik} (\dot q_i(\tau) + \dot N^m_{ik}(\tau)) - \nu_{ij}^m (\dot q_i(\tau) + \dot N^m_{ij}(\tau))
    \qs
    \label{eq:ref-sideband-fluctuations}
\end{align}
\end{subequations}

The \textit{\gls{tm} interferometer} also mixes the local and adjacent beams, and differs from the \gls{ref} interferometer only because the local beam is reflected off the \gls{tm}. Since we neglect any propagation inside the spacecraft, we simply have $\nu_{\text{tm}_{ij}} = \nu_{\text{ref}_{ij}}$.

\subsection{Measurement sampling}
\label{ssec:measurement-sampling}

The analog beatnotes are sampled, digitized and timestamped by the phasemeter, which uses a signal derived from the local \gls{uso} as timing reference~\cite{Barke:2015}. Therefore, all samples have timestamps attached that are equally sampled with respect to the local clock.

Formally, we need to express the signals we wish to sample as functions of the onboard clock times. For an analog signal $s(\tau)$ expressed in phase and as a function of the proper time, the same signal sampled to the clock time reads
\begin{equation}
    \tilde s(\tau) = s(\tau_i(\tau))
    \qc
\end{equation}
where $\tau_i(\tau)$ is the spacecraft proper time at which the onboard clock shows $\tau$. Here, $\tau_i(\tau)$ can be computed by inverting \cref{eq:clock-time-definition}. At first order in $q_i$, we have
\begin{equation}
    \tau_i(\tau) \approx \tau - q_i(\tau)
    \qs
\end{equation}
Using this result, we find at first order in $q_i$,
\begin{equation}
    \tilde s(\tau) \approx s(\tau) - \dot s(\tau) q_i(\tau)
    \qc
\end{equation}
which we can differentiate to obtain, in frequency,
\begin{equation}
\begin{split}
    \dv{\tilde{s}}{\tau} \qty(\tau) &\approx (1 - \dot q_i(\tau)) \dot s(\tau) - \ddot s(\tau) q_i(\tau)
    \\
    &\approx (1 - \dot q_i(\tau)) \dot s(\tau) 
    \qc
\end{split}
\label{eq:resampled-signal}
\end{equation}
where we neglect the term $\ddot s(\tau) q_i(\tau)$ since the total beatnote frequencies are evolving very slowly ($\ddot s(\tau) \approx \SI{E2}{\hertz\per\second}$)~\cite{Bayle:2021mue}.

Decomposing $\dot s(\tau)$ into frequency offsets $\dot s^o(\tau)$ and fluctuations $\dot s^\epsilon(\tau)$, and neglecting second-order terms in $\dot q_i(\tau) \dot s^\epsilon(\tau)$, \cref{eq:resampled-signal} becomes
\begin{equation}
    \dv{\tilde{s}}{\tau} \qty(\tau) \approx \dot s^o(\tau) - \dot q_i(\tau) \dot s^o(\tau) + \dot s^\epsilon(\tau)
    \qs
\end{equation}
We can now decompose again the resulting sampled signal into offsets and fluctuations,
\begin{subequations}
\begin{align}
    \dv{\tilde{s}^o}{\tau} \qty(\tau) &\approx \dot s^o(\tau)
    \qc
    \label{eq:resampled-offsets}
    \\
    \dv{\tilde{s}^\epsilon}{\tau} \qty(\tau) &\approx \dot s^\epsilon(\tau) - \dot q_i(\tau) \dot s^o(\tau)
    \qs
    \label{eq:resampled-fluctuations}
\end{align}
\end{subequations}

\subsection{Measurement equations}
\label{ssec:measurement-equations}

We saw in the previous section that clock noise only affects frequency fluctuations. Let us write the sampled beatnote frequency fluctuations from their analog descriptions given in the previous section. For concision, we will drop the tilde and denote $\text{isc}_{ij}(\tau)$, $\text{ref}_{ij}(\tau)$, and $\text{tm}_{ij}(\tau)$ the sampled \gls{isc}, \gls{ref}, and \gls{tm} beatnote frequency fluctuations, respectively. In addition, we will drop the explicit time dependency of all variables for the sake of readability, and introduce shortened notations for the time-varying beatnote frequency offsets,
\begin{equation}
    a_{ij} = \nu^o_{\text{isc}_{ij},\text{c}}
    \qand
    b_{ij} = \nu^o_{\text{ref}_{ij},\text{c}}
    \qs
\end{equation}

Plugging \cref{eq:isc-carrier-fluctuations,eq:isc-sideband-fluctuations} in \cref{eq:resampled-fluctuations} yields
\begin{subequations}
\begin{align}
    \text{isc}_{ij,\text{c}} &= - \dot q_i a_{ij}
    \qc
    \label{eq:sampled-isc-carrier-fluctuations}
    \\
    \begin{split}
        \text{isc}_{ij,\text{sb}} &= \nu^m_{ji} \ddelay{ij} (\dot q_j + \dot N^m_{ji}) - \nu_{ij}^m (\dot q_i + \dot N^m_{ij})
        \\
        &\qquad - \dot q_i (a_{ij} + \ddelay{ij} \nu^m_{ji} - \nu^m_{ij})
        \qs
    \end{split}
    \label{eq:sampled-isc-sideband-fluctuations}
\end{align}
\end{subequations}
Similarly, using \cref{eq:ref-carrier-fluctuations,eq:ref-sideband-fluctuations} gives
\begin{subequations}
\begin{align}
    \text{ref}_{ij,\text{c}} &= - \dot q_i b_{ij}
    \qc
    \label{eq:sampled-ref-carrier-fluctuations}
    \\
    \begin{split}
        \text{ref}_{ij,\text{sb}} &= \nu^m_{ik} (\dot q_i + \dot N^m_{ik}) - \nu_{ij}^m (\dot q_i + \dot N^m_{ij}) 
        \\
        &\qquad- \dot q_i(b_{ij} + \nu^m_{ik} - \nu^m_{ij})
        \qs
    \end{split}
    \label{eq:sampled-ref-sideband-fluctuations}
\end{align}
\end{subequations}
As discussed in \cref{ssec:interferometric-measurements}, \gls{tm} beatnote frequency fluctuations are identical to the \gls{ref} ones.

\section{Clock-noise reduction}
\label{sec:clock-noise-reduction}

\subsection{Detrending}
\label{ssec:detrending}

In reality, only the total beatnote frequencies (e.g., the total \gls{isc} carrier-carrier beatnote frequency $a_{ij} + \text{isc}_{ij,\text{c}}$) are directly measured by the phasemeter and telemetered to Earth. The current baseline for \gls{lisa} data processing is to filter out-of-band offsets from the total frequencies to recover the previously described in-band frequency fluctuations ($\text{isc}_{ij,\text{c}}$ in the previous example).

In the following, we assume that this process can be achieved without errors, and will use the beatnote frequency fluctuations described in \cref{ssec:measurement-equations}.

\subsection{Intermediary variables}

The so-called \gls{tdi} intermediary variables $\xi$ are constructed to remove spacecraft jitter noise~\cite{Otto:2015wp},
\begin{subequations}
\begin{equation}
    \xi_{ij} = \text{isc}_{ij,\text{c}} + \frac{(\text{ref}_{ij,\text{c}} - \text{tm}_{ij,\text{c}})}{2} + \ddelay{ij} \frac{(\text{ref}_{ji,\text{c}} - \text{tm}_{ji,\text{c}})}{2}
    \qs
\end{equation}
\end{subequations}
Next up, the $\eta$ intermediary variables remove half of the laser noise, respectively. They read~\cite{Otto:2015wp},
\begin{subequations}
\begin{align}
    \eta_{ij} &= \xi_{ij} + \ddelay{ij} \frac{\text{ref}_{ji,\text{c}} - \text{ref}_{jk,\text{c}}}{2}
    \qc
    \\
    \eta_{ik} &= \xi_{ik} + \frac{\text{ref}_{ij,\text{c}} - \text{ref}_{ik,\text{c}}}{2}
    \qc
\end{align}
\end{subequations}
for left $ij$ and right $ik$ \glspl{mosa}, respectively.

Inserting \cref{eq:sampled-isc-carrier-fluctuations,eq:sampled-ref-carrier-fluctuations} in these expressions, we find how clock noise enters these intermediary variables,
\begin{align}
    \eta_{ij} &= \ddelay{ij} b_{jk} \dot q_j - a_{ij} \dot q_i
    \qc
    \label{eq:q-in-eta-left}
    \\
    \eta_{ik} &= -(b_{ij} + a_{ik}) \dot q_i
    \qs
    \label{eq:q-in-eta-right}
\end{align}
Note that $b_{ij} = -b_{ik}$. In the following, we choose to only use reference beatnote frequencies $b_{ij}$ from left \glspl{mosa}.

\subsection{Clock-noise residuals}

From these intermediary variables, we can build laser noise-free \gls{tdi} combinations. They can be expressed as polynomials of delay operators $\poly{ij}$, in the form
\begin{equation}
    \text{TDI} = \sum_{i,j \in \idxset_2} \poly{ij} \eta_{ij}
    \qs
    \label{eq:general-tdi}
\end{equation}
where $\idxset_2 = \qty{\qty(1,2), \qty(2,3), \qty(3,1), \qty(1,3), \qty(3,2), \qty(2,1)}$ is the set of the 6 \gls{mosa} index pairs.

Inserting \cref{eq:q-in-eta-left,eq:q-in-eta-right}, we find that clock noise enters in the \gls{tdi} combination as
\begin{equation}
    \text{TDI}^q = \sum_{i,j,k \in \idxset_3^+}{[\poly{ki} \ddelay{ki} - \poly{ik}] b_{ij} \dot q_i} - \sum_{i,j \in \idxset_2}{\poly{ij}a_{ij} \dot q_i}
    \qc
    \label{eq:q-in-tdi}
\end{equation}
with $\idxset_3^+ = \qty{\qty(1,2,3),\qty(2,3,1),\qty(3,1,2)}$ as the set of triplets of spacecraft indices in ascending order.

To estimate the contribution of clock noise before any correction, we assume that all light travel times are constant and equal to $L$, such that we can commute delay operators. As shown in \cite{Bayle:2019}, these commutators only yield multiplicative terms~$\ll 1$. Also, we suppose that all clock noises are uncorrelated but have the same \gls{psd} $\psd{\dot q}(\omega)$. Lastly, we assume that beatnote frequency offsets are constant. The clock noise residual \gls{psd} then reads
\begin{equation}
\begin{split}
    &\psd{\text{TDI}^q}(\omega) \approx \sum_{i,j,k \in \idxset_3^+} \Big|a_{ij} \fourierpoly{ij}(\omega)+ a_{ji}  \fourierpoly{ji}(\omega)
    \\
    &\qquad -b_{ij} [\fourierpoly{ik}(\omega) - \fourierpoly{ki}(\omega) \fourierddelay{ki}(\omega)] \Big|^2 \psd{\dot q}(\omega)
    \qc
    \label{eq:tdi-clock-residuals-psd}
\end{split}
\end{equation}

Here, $\fourierddelay{ij}$ and $\fourierpoly{ij}$ are the Fourier transforms of delay operators and polynomials thereof, see~\cite{Bayle:2019} for further information.

As an example, we can use \cref{eq:tdi-clock-residuals-psd} to work out clock noise residuals in the second-generation Michelson combination $X_2$, as defined in \cite{Otto:2015wp},
\begin{equation}
\begin{split}
    &X_2 = (1 - \ddelay{121} - \ddelay{12131} + \ddelay{1312121}) (\eta_{13} + \ddelay{13} \eta_{31}) \\
    &\quad - (1 - \ddelay{131} - \ddelay{13121} + \ddelay{1213131}) (\eta_{12} + \ddelay{12} \eta_{21})
    \qs
    \label{eq:X2}
\end{split}
\end{equation}
We find that
\begin{equation}
    \psd{X_2^q}(\omega) \approx 16 \sin[2](2 \omega L) \sin[2](\omega L) A_{X_2}(\omega) \psd{\dot q}(\omega)
    \qc
    \label{eq:q-in-X2-psd}
\end{equation}
with $A_{X_2}(\omega)$ is a scaling factor that depends only on the beatnote frequencies,
\begin{equation}
\begin{split}
    &A_{X_2}(\omega) = (a_{12} - a_{13})^2 + a_{21}^2 + a_{31}^2
    \\
    &\qquad - 4 b_{12} (a_{12} - a_{13} - b_{12}) \sin[2](\omega L)
    \qs
    \label{eq:A-X2-definition}
\end{split}
\end{equation}

\begin{figure}
    \centering
    \includegraphics[width=\columnwidth]{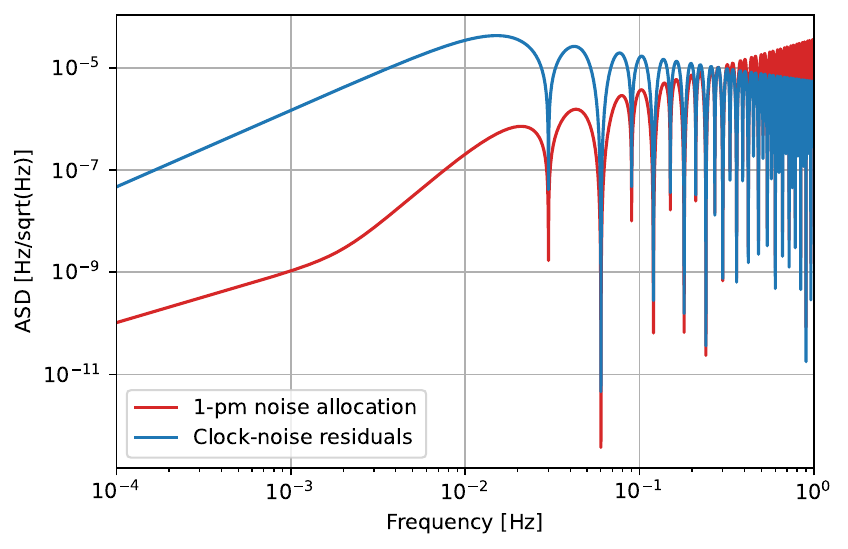}
    \caption{Comparison of the residual clock noise in second-generation Michelson $X_2$ combination, and the usual \gls{lisa} \SI{1}{\pico\meter} noise allocation curve. We assumed here a state-of-the-art space-qualified \gls{uso} and a realistic set of beatnote frequency offsets.}
    \label{fig:q-in-X2}
\end{figure}

\Cref{fig:q-in-X2} shows this clock noise residual for a state-of-the-art space-qualified \gls{uso} with $\psd{\dot q}(f) = \num{4E-27} f^{-1}$ in fractional frequency fluctuations~\cite{jpl-talk}, and a realistic set of beatnote frequency offsets. We compared it to a typical \SI{1}{\pico\meter} \gls{lisa} noise allocation for a single noise source,
\begin{equation}
\begin{split}
    &\psd{X_2^\text{alloc}}(\omega) = 64 \omega^2 \sin[2](\omega L) \sin[2](2 \omega L)  \\
    &\qquad \times \qty(\frac{\SI{1}{\pico\meter\hertz^{-1/2}}}{\lambda})^2 \qty[1 + \qty(\frac{\SI{2e-3}{\hertz}}{\omega / 2 \pi})^4]
    \qs
    \label{eq:noise-allocation}
\end{split}
\end{equation}
We see that below \SI{0.2}{\hertz}, clock noise violates this requirement and must be suppressed.

\subsection{Building correcting expression}

Inspecting \cref{eq:q-in-tdi}, we observe that clock noise enters in our \gls{tdi} combination coupled with delay polynomials. We can rearrange it to get
\begin{equation}
\begin{split}
    \text{TDI}^q =  \sum_{i,j,k \in \idxset_3^+} & \poly{ij}  (b_{jk} - a_{ij}) \dot q_i
    \\
    &\quad - \poly{ik} (b_{ij} + a_{ik}) \dot q_i \\
    &\quad + \poly{ij} (\ddelay{ij} b_{jk} \dot q_j - b_{jk} \dot q_i)
    \qs
    \label{eq:q-in-tdi-rearranged}
\end{split}
\end{equation}

In the previous equations, beatnote frequency offsets $a_{ij},\  b_{ij}$ are time-dependent, c.f.~\cref{eq:isc-carrier-offsets}. As such, commuting them with a delay operator yields an error term. This effect will be studied in \cref{ssec:time-varying-beatnote-frequencies}. In this section, we neglect it and write the previous equation as
\begin{equation}
\begin{split}
    \text{TDI}^q \approx \sum_{i,j,k \in \idxset_3^+} &  (b_{jk} - a_{ij}) \poly{ij} \dot q_i
    \\
    &\quad - (b_{ij} + a_{ik}) \poly{ik} \dot q_i \\
    &\quad + b_{jk} \poly{ij} (\ddelay{ij} \dot q_j - \dot q_i)
    \qs
    \label{eq:q-in-tdi-commuted}
\end{split}
\end{equation}

We observe that clock noise appears on the right side of the $\poly{ij}$, either \textit{directly} as $\dot q_i$, or as a \textit{differential term} $\ddelay{ij} \dot q_j - \dot q_i$.

We will now see that the sideband-sideband beatnotes introduced in \cref{ssec:interferometric-measurements} can provide a direct measurement of the differential clock noise. We form the dimensionless quantity
\begin{equation}
    r_{ij} = \frac{\text{isc}_{ij,\text{sb}} - \text{isc}_{ij,\text{c}} }{\nu^m_{ji}}
    \qs
    \label{eq:r-definition}
\end{equation}
By inserting \cref{eq:sampled-isc-carrier-fluctuations,eq:sampled-isc-sideband-fluctuations} into the previous expression, we see that $r_{ij}$ directly measures the differential clock noise appearing in \cref{eq:q-in-tdi-rearranged},
\begin{equation}
    r_{ij} = \ddelay{ij} \dot q_j - \dot q_i
    \qc
    \label{eq:q-in-r}
\end{equation}
where we neglect the modulation noise for now. The effect of modulation noise and an additional processing step for its partial mitigation are discussed in \cref{ssec:modulation-errors}.

Because the beatnote frequency offsets are measured and the delay polynomials are known, we can  subtract $\sum_{i,j,k \in \idxset_3^+}{b_{jk} \poly{ij} r_{ij}}$ from our \gls{tdi} combination to remove the last term of \cref{eq:q-in-tdi-rearranged}.

We are left with the first two terms in the sum in \cref{eq:q-in-tdi-rearranged}. Vallisneri discusses in~\cite{Vallisneri:2005} how any arbitrary interferometer can be synthesized out of a set of 6 one-way interferometric measurements $y_{ij}$ between the \gls{lisa} spacecraft. The $r_{ij}$ are of exactly the same form as the $y_{ij}$ used in \cite{Vallisneri:2005}, just expressed in terms of clock noise instead of laser noise. Therefore, we can apply the same algorithm\footnote{Note that \cite{Vallisneri:2005} uses a different notation. We translated their algorithm to our notation in \cref{sec:geometric-tdi-algorithm}.} to construct any expression of the form $(\ddelay{i_1, i_2, \dots, i_n} q_{i_n} - q_{i_1})$ from the $r_{ij}$. 

Each delay polynomial $\poly{ij}$ in \cref{eq:general-tdi} is the sum of $n_{ij}$ chained delay operators with signs $\sigma_{ij}^k$. It can be written as
\begin{equation}
    \poly{ij} = \sum_{k=1}^{n_{ij}}{\sigma_{ij}^k \ddelay{A_{ij}^k, \dots, B_{ij}^k}}
    \qc
\end{equation}
where we denote the first and last indices of the chained delay operators in the $k$-th summand with $A_{ij}^k$ and $B_{ij}^k$, respectively. By design of the \gls{tdi} algorithm, we always have $B_{ij}^k = i$.

Using Vallisneri's algorithm for each summand, we can construct
\begin{equation}
    R_{ij} = \sum_{k=1}^{n_{ij}}{\sigma_{ij}^k \qty(\ddelay{A_{ij}^k, \dots, i} \dot q_{i} - \dot q_{A_{ij}^k})}
    \qs
    \label{eq:definition-Rij}
\end{equation}

We have
\begin{equation}
    \poly{ij} \dot q_i - R_{ij} = \sum_{k=1}^{n_{ij}}{\sigma_{ij}^k \dot q_{A_{ij}^k}}
    \qs
    \label{eq:Rij-condition}
\end{equation}
In most second generation \gls{tdi} combinations\footnote{We explicitly tested it for all second generation \gls{tdi} combinations up to 16 links presented in \cite{Muratore:2020mdf}. We conjecture that it is valid for any second generation variable derived from geometric principles.}, this last term is vanishing, such that we have $R_{ij} = P_{ij} q_i$. Therefore, we can subtract them from \cref{eq:q-in-tdi-rearranged} to remove the last clock-noise terms.

In the special case where one summand in $\poly{ij}$, say the $n$-th term, does not contain any delay, we skip it in the construction of $R_{ij}$ (\cref{eq:definition-Rij}). An extra term $\sigma_{ij}^n \dot q_i$ is to be accounted for in \cref{eq:Rij-condition}, which has to cancel with one of the $\sigma_{ij}^k \dot q_{A_{ij}^k}$.

The full corrected \gls{tdi} combination therefore reads
\begin{equation}
\begin{split}
    \text{TDI}^c = \text{TDI} - \sum_{i,j,k \in \idxset_3^+} &(b_{jk} - a_{ij}) R_{ij}
    \\
    &\quad - (b_{ij} + a_{ik}) R_{ik} \\
    &\quad + b_{jk} \poly{ij} r_{ij}
    \qs
\end{split}
\label{eq:corrected-tdi}
\end{equation}

Note that this algorithm can be applied without modification to \gls{tdi} variables containing not only delays, but also advancements~\cite{Vallisneri:2005,Muratore:2020mdf}.

\subsection{Example: correcting clock noise for \texorpdfstring{$X_2$}{X2}}

As an example, we apply our algorithm to the second-generation Michelson combination $X_2$ given in \cref{eq:X2}. Following the decomposition of \cref{eq:general-tdi},
\begin{subequations}
\begin{align}
    \poly{12} &= - (1 - \ddelay{131} - \ddelay{13121} + \ddelay{1213131})
    \qc
    \\
    \poly{23} &= 0
    \qc
    \\
    \poly{31} &= (1 - \ddelay{121} - \ddelay{12131} + \ddelay{1312121}) \ddelay{13}
    \qc
    \\
    \poly{21} &= - (1 - \ddelay{131} - \ddelay{13121} + \ddelay{1213131}) \ddelay{12}
    \qc
    \\
    \poly{32} &= 0
    \qc
    \\
    \poly{13} &= (1 - \ddelay{121} - \ddelay{12131} + \ddelay{1312121})
    \qs
\end{align}
\end{subequations}

Applying Vallisneri's algorithm, we construct the $R_{ij}$ variables verifying \cref{eq:definition-Rij}. They read
\begin{subequations}
\begin{align}
    \begin{split}
        R_{12} &= - (1 - \ddelay{131}) (r_{12} + \ddelay{12} r_{21}) 
        \\
        &\qquad + (2 - \ddelay{121} - \ddelay{12131}) (r_{13} + \ddelay{13} r_{31}) 
        \qc
    \end{split}
    \\
    R_{23} &= 0
    \qc
    \\
    \begin{split}
        R_{31} &= - (2 - \ddelay{131} - \ddelay{13121}) (r_{12} + \ddelay{12} r_{21})
        \\
        &\qquad + (1 - \ddelay{121})(r_{13} + \ddelay{13} r_{31})
        \\
        &\qquad + (1 - \ddelay{121} - \ddelay{12131} + \ddelay{1312121}) r_{13}
        \qc
    \end{split}
    \\
    \begin{split}
        R_{21} &= - (1 - \ddelay{131}) (r_{12} + \ddelay{12} r_{21})
        \\
        &\qquad - (1 - \ddelay{131} - \ddelay{13121} + \ddelay{1213131}) r_{12}
        \\
        &\qquad + (2 - \ddelay{121} - \ddelay{12131}) (r_{13} + \ddelay{13} r_{31})
        \qc
    \end{split}
    \\
    R_{32} &= 0
    \qc
    \\
    \begin{split}
        R_{13} &= - (2 - \ddelay{131} - \ddelay{13121}) (r_{12} + \ddelay{12} r_{21})
        \\
        &\qquad + (1 - \ddelay{121}) (r_{13} + \ddelay{13} r_{31})
        \qs
    \end{split}
\end{align}
\end{subequations}
These can be directly inserted into \cref{eq:corrected-tdi} to get the corrected variable $X_2^c$.


\section{Numerical simulations}
\label{sec:simulations}

To complement the theoretical studies presented above, we conducted numerical experiments to verify the clock-noise suppression capabilities of this algorithm. We simulated the signals presented in \cref{ssec:interferometric-measurements} using the Python tool LISA Instrument, which is based on the official LISA Consortium simulator LISANode~\cite{Bayle:2019dfu}. We then used the Python tool PyTDI to generate the second-generation Michelson TDI data streams $X_2$, as well as the corresponding clock correction, all described in \cref{sec:clock-noise-reduction}.

Laser beams and interferometric measurements are simulated in frequency. In addition, we use the decomposition in frequency offsets and fluctuations presented in \cref{ssec:description-of-laser-beams}. Each of these variables is represented by a 64-bit floating-point number (double precision), such that they have sufficient dynamic range not to be limited by quantization noise. Note that in reality, we will only have access to the sum of these two variables, and need to decompose them on-ground before we can apply this clock-correction algorithm (c.f. \cref{ssec:detrending}).

In order not to be limited by other numerical effects, such as interpolation errors or flexing-filtering couplings~\cite{Bayle:2019}, we use an unrealistically high sampling rate of \SI{10}{\hertz} for the full simulation chain. By doing so, we omit any simulation of on-board filters. The duration of our simulation is set to \SI{E6}{\second}, i.e., \num{E7} samples.

The propagation of laser beams between spacecraft is implemented using Lagrange fractional delay filters of order 31. Given our high sampling rate, they do not cause any observable artifact in our frequency band of interest and remain computationally tractable. Light travel times and their derivatives are computed from realistic orbits provided by \gls{esa}. These light travel times include relativistic corrections up to terms in $1/c$, including the Sagnac effect and the Shapiro delay.

We use the following programmed offsets, see \cref{eq:local-carrier-offsets},
\begin{align}
    O_{12} &= \SI{8.1}{\mega\hertz}
    \qcomma
    O_{21} = \SI{-9.5}{\mega\hertz}
    \qc
    \\
    O_{13} &= \SI{1.4}{\mega\hertz}
    \qcomma
    O_{31} = \SI{10.3}{\mega\hertz}
    \qc
    \\
    O_{23} &= \SI{9.2}{\mega\hertz}
    \qcomma
    O_{32} = \SI{-11.6}{\mega\hertz}
    \qs
\end{align}

Our simulated clock noise matches that of a state-of-the-art space-qualified \gls{uso} with
\begin{equation}
    \psd{\dot q}(f) = \num{4E-27} f^{-1}
    \label{eq:clock-noise-psd}
\end{equation}
in fractional frequency deviations, generated using a variant of the infinite RC model~\cite{Plaszczynski:2005yp}. We do not simulate any deterministic clock errors, such as constant timing and frequency offsets or higher order frequency drifts due to aging of the oscillators.
    
Modulation errors are based on a fit of the \SI{2}{\watt}-fiber amplifier (red) curve from figure 5.13 in~\cite{Barke:2015}. In fractional frequency deviations with respect to the modulation frequency, they are given by
\begin{equation}
    \psd{\dot N^m}(f) = \num{5.2E-14} f^{1/3}
    \qs
    \label{eq:modulation-noise-level}
\end{equation}
Note that this noise level does not meet the requirements for \gls{lisa}~\cite{Barke:2015}, and consequently also violates our \SI{1}{\pico\meter}-noise allocation curve. Laser development is ongoing to achieve the required noise levels in line with the stringent timing requirements\footnote{M. Hewitson, private communication, (April 9, 2021).}.

Because of \gls{lisa}'s frequency distribution system design~\cite{Barke:2015}, only one of the two modulation signals on a spacecraft has a direct low-noise relationship to the pilot tone used as timing reference. Therefore, to demonstrate the correctness of our algorithm, we make the working assumption that the right hand-side modulation noises (indexed 13, 32, 21) are 10 times higher, i.e. with a spectrum 
\begin{equation}
    \psd{\dot N^m}(f) = \num{5.2E-13} f^{1/3}
    \qs
    \label{eq:right-modulation-noise-level}
\end{equation}

To compute our \gls{tdi} combinations, we also use order-31 Lagrange fractional delay filters.

\begin{figure*}
    \centering
    \includegraphics[width=\textwidth]{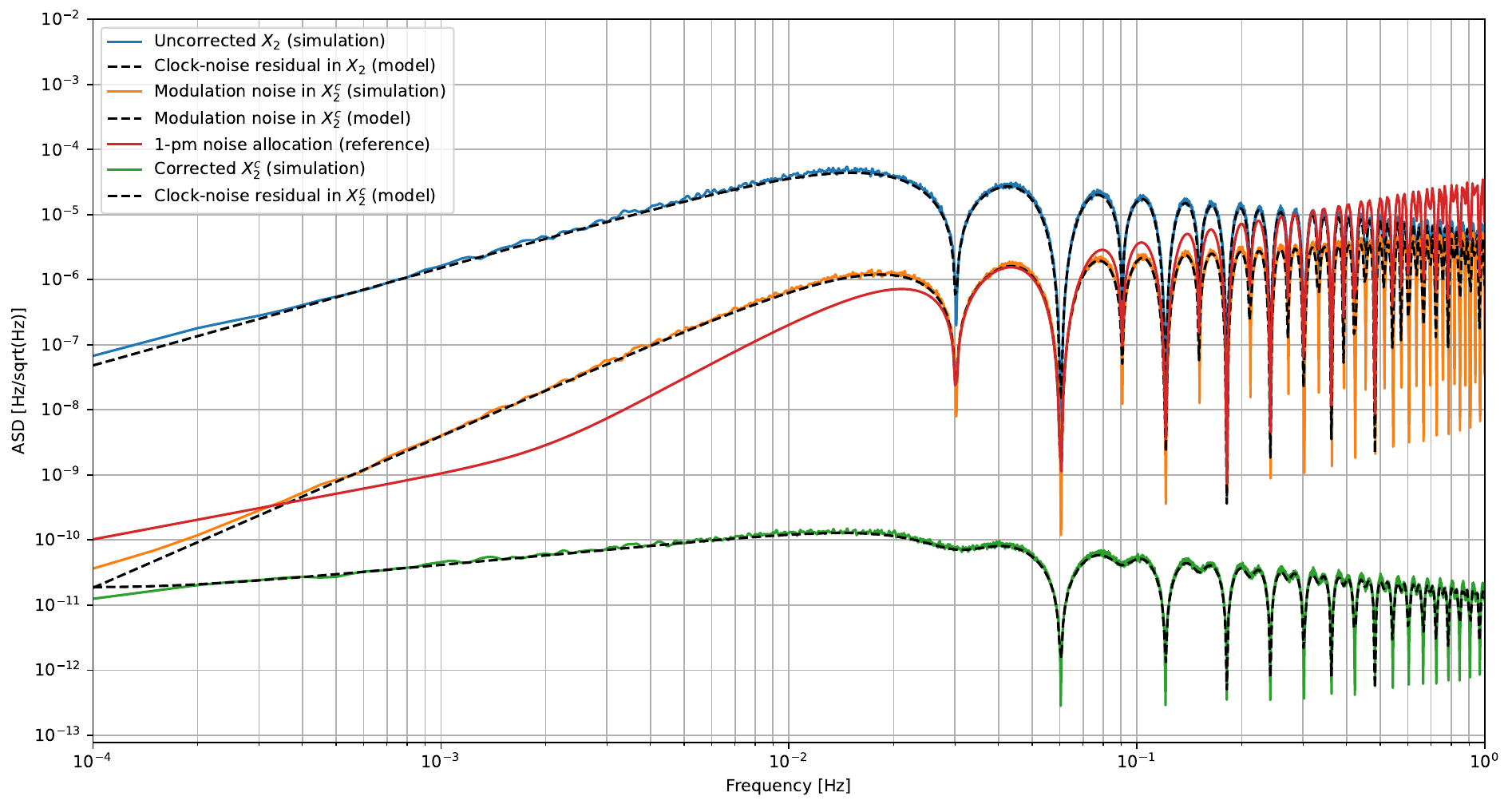}
    \caption{Simulation results. Blue and green curves represent the uncorrected $X_2$ and corrected $X_2^c$ combinations in the sole presence of clock noise. The yellow curve shows the level of modulation noise in the corrected variable. Overlaid dashed black lines show our analytical expectations for these quantities. The usual \SI{1}{\pico\meter}-noise allocation curve is shown in red as a reference.}
    \label{fig:simulation}
\end{figure*}

\Cref{fig:simulation} presents the results of our numerical experiments. We used Welch method to estimate the spectra, with segments of \num{200000} samples and a \textit{Nutall4} window function. As a reference, we also plot in red the usual \gls{lisa} Performance Model's \SI{1}{\pico\meter}-noise allocation curve, given by \cref{eq:noise-allocation}.

The blue and green curves represent the \glspl{asd} of the uncorrected $X_2$ and corrected $X_2^c$ second-generation Michelson \gls{tdi} combinations, in the sole presence of clock noise (we disabled modulation errors). The black dashed line overlaying the former curve represents the theoretical clock-noise content in $X_2$ from \cref{eq:q-in-X2-psd}. The black dashed line overlaying the green curve, representing our model of the residual clock noise after correction, is discussed in details in \cref{ssec:time-varying-beatnote-frequencies} and given by \cref{eq:time-varying-beatnote-residual}.

The orange curve represents the modulation errors in the corrected $X_2^c$ combination. We obtain it by enabling modulation errors in addition to clock noise in the previous simulation. In addition, we perform an additional processing step to remove the higher right hand-side modulation noise terms, see \cref{ssec:modulation-errors}. The overlaid black dashed line shows our analytical expectations, also described in the same section.

\section{Discussion}
\label{sec:discussion}

We see in \cref{fig:simulation} that our clock-noise reduction algorithm works as expected and that clock noise is reduced to below the noise allocation curve. In addition, our analytical models match perfectly with simulated results.

We discuss in the following paragraphs the models for the residual clock noise after correction, the imperfections in the sideband measurements used for the correction, and other limiting effects that may appear with different simulation parameters but were not visible in our setup. Finally, we compare our algorithm with the previously proposed solution~\cite{Tinto:2018}.

\subsection{Time-varying beatnote frequencies}
\label{ssec:time-varying-beatnote-frequencies}

The residual clock noise after correction is dominated by the effect of time-varying beatnote frequencies, which we neglected when deriving the correction. To evaluate this clock-noise residuals, we take the difference between \cref{eq:q-in-tdi-rearranged} and \cref{eq:q-in-tdi-commuted}. One gets
\begin{equation}
    \text{TDI}^{\nu} = \sum_{i,j,k \in \idxset_3^+}{\qty(\comm{a_{ij}}{\poly{ij}} + \comm{a_{ik}}{\poly{ik}}) q_i}
    \qc
\end{equation}
where we used that the reference beatnote offsets are constants in our setup, see \cref{eq:ref-carrier-offsets}, and hence can be freely commuted with delays. Here, $\comm{A}{B} = AB - BA$ stands for the commutator of $A$ and $B$.

We can now assume that all delays are equal and constant, such that a delay operator of $N$ delays can be written as $\delay{}^{N}$. We can write the delay polynomial $\poly{ij}$ as
\begin{equation}
    \poly{ij} = \sum_\alpha{\lambda_\alpha \delay{}^{N_\alpha}}
    \qc
\end{equation}
with $\lambda_\alpha$ the factor in front of the term $\alpha$, and $N_\alpha$ the number of delays associated with the same term. We now have
\begin{equation}
    \comm{a_{ij}}{\poly{ij}} = \sum_\alpha{\lambda_\alpha \comm{a_{ij}}{\delay{}^{N_\alpha}}}
    \qc
\end{equation}
and similarly for $\poly{ik}$.

Let us study one commutator $\comm{a_{ij}}{\delay{}^N}$. This is an operator, which will be applied on $q_i$. 
Assuming that $a_{ij}$ is a linear function of time, we can write
\begin{align*}
    \comm{a_{ij}}{\delay{}^N} &= a_{ij}(t) \delay{}^N - (\delay{}^N a_{ij}) \delay{}^N
    \\
    &= (a_{ij} - \delay{}^N a_{ij}) \delay{}^N
    \\
    &= N d\, \dot a_{ij} \delay{}^N
    \qc
\end{align*}
where $N d\, \dot a_{ij}$ is a constant. We can then go to the frequency domain to write it as
\begin{equation}
    N d\, \dot a_{ij} e^{-j \omega N d}
    \qs
\end{equation}

We can therefore define
\begin{equation}
    Q_{ij}(\omega) = \sum_\alpha{\lambda_\alpha  N_\alpha d\, e^{-j \omega d N}}
    \qc
\end{equation}
such that, still in the frequency domain,
\begin{equation}
    \comm{a_{ij}}{\poly{ij}} = Q_{ij} \dot a_{ij}
    \qs
\end{equation}
Finally, using that clocks are independent but share the same statistical properties, we obtain the following residual from the time-varying beatnote offsets,
\begin{equation}
    \psd{\text{TDI}^{\nu}}(\omega) = \psd{q}(\omega) \sum_{i,j,k \in \idxset_3^+}{\abs{Q_{ij}(\omega) \dot a_{ij} + Q_{ik}(\omega) \dot a_{ik}}^2}
    \qs
    \label{eq:time-varying-beatnote-residual}
\end{equation}

In our simulation, the average value of the beatnote derivatives evaluate to
\begin{align}
    \dot a_{12} &\approx 
    \dot a_{21} \approx \SI{-2.0}{\hertz\per\second}
    \qc
    \\
    \dot a_{13} &\approx
    \dot a_{31} \approx \SI{320}{\milli\hertz\per\second}
    \qc
    \\
    \dot a_{23} &\approx
    \dot a_{32} \approx \SI{-58}{\milli\hertz\per\second}
    \qc
\end{align}
while their average second derivatives are of the order of, or below, \SI{E-7}{\hertz\per\second\squared}, which is small enough to neglect it in our model. This is also verified by the perfect agreement between model and simulation in \cref{fig:simulation}.

\subsection{Modulation errors}
\label{ssec:modulation-errors}

In \cref{ssec:description-of-laser-beams}, we introduce modulation errors to model imperfections of the sideband modulation. By inserting \cref{eq:sampled-isc-carrier-fluctuations,eq:sampled-isc-sideband-fluctuations} into \cref{eq:r-definition}, we see that they appear in $r_{ij}$ as
\begin{equation}
    r_{ij} = \frac{\nu^m_{ji} \ddelay{ij} \dot N^m_{ji} - \nu_{ij}^m \dot N^m_{ij}}{\nu^m_{ji}}
    \qs
\end{equation}
Therefore, our clock noise reduction performance will ultimately be limited by this modulation noise. To achieve the required performance, we must remove the higher modulation noise contributions from the right hand-sided modulation signals.

To remove these higher modulation noise terms, we need to measure the difference $\Delta M_i$ of the two modulation signals on each spacecraft $i$. It can be measured electrically before the optical modulations are performed, or optically, using the sideband-sideband beatnotes in the reference interferometers.

We study here the second option, and define
\begin{align}
    \Delta M_i &=
    \frac{\text{ref}_{ik, \text{sb}} - \text{ref}_{ik}}{2} - \frac{\text{ref}_{ij, \text{sb}} - \text{ref}_{ij}}{2}
    \qc
    \label{eq:definition-delta-M}
\end{align}
where the indices $i,j,k \in \idxset_3^+$. 

These expressions are free of laser noise. Inserting \cref{eq:sampled-ref-carrier-fluctuations,eq:sampled-ref-sideband-fluctuations}, we see that they contain the modulation noise terms
\begin{align}
    \Delta M_i &\approx \nu^m_{ij} N^m_{ij} - \nu^m_{ik} N^m_{ik}
    \qs
\end{align}
Note that both terms in \cref{eq:definition-delta-M} contain the same information. We use here interferometric measurements from both adjacent \glspl{mosa} $ij$ and $ik$ to reduce the overall readout noise in $\Delta M_i$, which is uncorrelated in both measurements.

We now define
\begin{subequations}
\begin{align}
    r^c_{ij} &= r_{ij} + \frac{\ddelay{ij} \Delta M_j}{\nu^m_{ji}} 
    \qc 
    \\
    r^c_{ik} &= r_{ik} - \frac{\Delta M_i}{ \nu^m_{ki}}
    \qc
\end{align}
\end{subequations}
which contain the same differential clock noise as $r_{ij}$ and $r_{ik}$, so that we can use them in place of the latter in the clock-noise reduction procedure described in \cref{sec:clock-noise-reduction}. However, substituting previous expressions, it becomes clear that these terms do not contain modulation noise from right-handed \glspl{mosa}, as we have
\begin{subequations}
\begin{align}
    r^c_{ij} &= \frac{\ddelay{ij} \nu^m_{jk} N^m_{jk} - \nu^m_{ij} N^m_{ij}}{\nu^m_{ji}}
    \qc 
    \\
    r^c_{ik} &= \frac{\ddelay{ik} \nu^m_{ki} N^m_{ki} - \nu^m_{ij} N^m_{ij}}{\nu^m_{ki}}
    \qs
\end{align}
\end{subequations}

Since we choose $\nu^m_{ij}$ identical on on all left-handed \glspl{mosa}, this simplifies to
\begin{subequations}
\begin{align}
    r^c_{ij} &= \ddelay{ij} N^m_{jk} - N^m_{ij}
    \qc \label{eq:rijc-residual}
    \\
    r^c_{ik} &= \ddelay{ik}  N^m_{ki} -  N^m_{ij}
    \qc \label{eq:rikc-residual}
\end{align}
\end{subequations}

The remaining left-sided modulation error terms enter \cref{eq:rijc-residual,eq:rikc-residual} with the same pattern as clock noise in \cref{eq:q-in-r}. This extends to the clock noise correction term we subtract in \cref{eq:corrected-tdi}, such that the overall modulation noise exactly replaces the clock noise in the original \gls{tdi} expression. As such, following \cref{eq:tdi-clock-residuals-psd}, their \glspl{psd} are given by
\begin{equation}
\begin{split}
    &\sum_{i,j,k \in \idxset_3^+} \Big|a_{ij} \fourierpoly{ij}(\omega)+ a_{ji} \fourierpoly{ji}(\omega)
    \\
    &\qquad -b_{ij} [\fourierpoly{ik}(\omega) - \fourierpoly{ki}(\omega) \fourierddelay{ki}(\omega)] \Big|^2 \psd{\dot N^m}(\omega)
    \qc
    \label{eq:tdi-modulation-residuals-psd}
\end{split}
\end{equation}
where we assume that modulation noise is uncorrelated but of equal \gls{psd} $\psd{N^m}(\omega)$ for all \SI{2.4}{\giga\hertz} sidebands.

In the particular case of $X_2$, we get
\begin{equation}
    16 \sin[2](2 \omega L) \sin[2](\omega L) A_{X_2}(\omega) \psd{\dot N^m}(\omega)
    \qc
\end{equation}
with $A_{X_2}$ as the same scaling factor as given in \cref{eq:A-X2-definition},
\begin{equation}
\begin{split}
    &A_{X_2}(\omega) = (a_{12} - a_{13})^2 + a_{21}^2 + a_{31}^2
    \\
    &\qquad - 4 b_{12} (a_{12} - a_{13} - b_{12}) \sin[2](\omega L)
    \qs
\end{split}
\end{equation}
We see in \cref{fig:simulation} that this model fits perfectly our numerical results.

\subsection{Calibration errors}

The clock correction algorithm relies on accurate measurements of the large beatnote frequency offsets $a_{ij}, b_{ij}$. Therefore, the achievable clock correction will be limited to a level proportional to the fractional error in these coefficients. A simple model for the residual noise from this effect can be derived by replacing the coefficients $a_{ij}, b_{ij}$ in \cref{eq:q-in-X2-psd} with terms representing the calibration error in the respective variables.

In reality, $a_{ij}, b_{ij}$ will be determined with respect to the frequency of the \gls{uso}, such that large deterministic offsets in its frequency will affect their measured value. We neglected such offsets in our analytical study and simulation, but remark that they are typically at a level below \num{e-6} for a space-qualified \gls{uso}~\cite{jpl-talk}, such that the residual noise due to this effect is smaller then the one presented in \cref{ssec:time-varying-beatnote-frequencies}.

\subsection{Other effects}

The clock-noise reduction algorithm presented here relies on the same principles as the laser-noise reduction by the traditional \gls{tdi} combinations. As such, any effect which limits the latter will also impact the former in a similar fashion. Examples of such effects include the flexing-filtering coupling discussed in \cite{Bayle:2019}, aliasing of power due to insufficient filtering at high frequencies, interpolation artifacts around the Nyquist frequency, or the so-called \textit{ranging errors} due to uncertainties in the values of the delays to apply.

However, as seen in \cref{fig:simulation}, clock noise only needs to be reduced by around 4 orders of magnitude, whereas laser noise needs to be reduced by around 8 orders of magnitude\footnote{For a laser noise of \SI{30}{\hertz\hertz^{-1/2}} compared to the usual \gls{lisa} \SI{1}{\pico\meter} noise allocation curve.}. Therefore, if these effects are sufficiently small to allow for laser noise reduction, they should automatically be sufficiently small to not impact the clock noise reduction.

\subsection{Comparison to other algorithms}
\label{ssec:comparison-to-other-algorithms}

\begin{figure}
    \centering
    \includegraphics[width=\columnwidth]{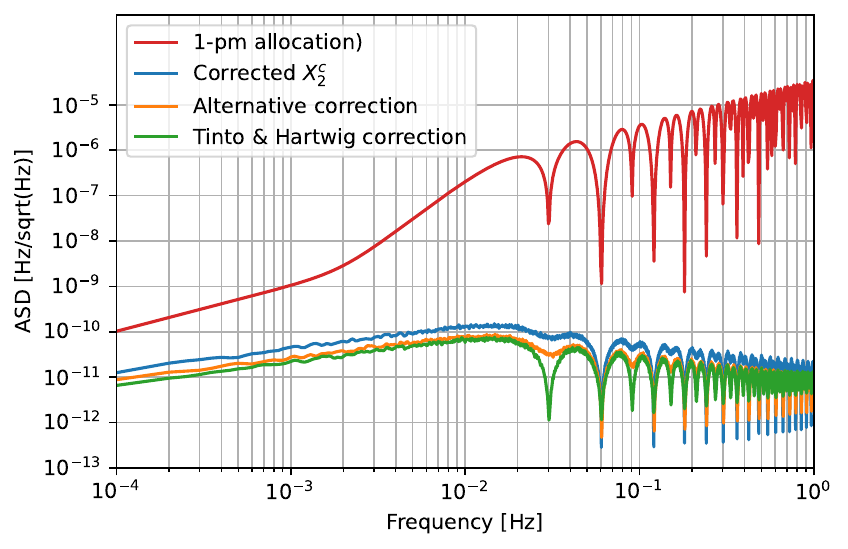}
    \caption{Comparison of the residual clock noise obtained with the reduction algorithm proposed in \cref{sec:clock-noise-reduction} (in blue), Tinto \& Hartwig's algorithm from~\cite{Tinto:2018} (in yellow), and the alternative algorithm proposed in \cref{ssec:comparison-to-other-algorithms} (in green). The usual \SI{1}{\pico\meter}-noise allocation curve is shown in red.}
    \label{fig:alternative}
\end{figure}

In \cref{fig:alternative}, we compare the performance of our clock-noise reduction algorithm (blue curve) against that from~\cite{Tinto:2018} (plotted in yellow). In our notation system and replacing delay operators by their Doppler-delay counterparts, the former algorithm is given by
\begin{equation}
    X_2^c = X_2 - (1 - \ddelay{12131}) K_{X_1}
    \qc
\end{equation}
where $K_{X_1}$ is the correcting expression for the first-generation Michelson,
\begin{equation}
\begin{split}
    &K_{X_1} = -\frac{b_{12}}{2}
    \Big[ (1 - \ddelay{121})(r^c_{13} + \ddelay{21} r^c_{31})
    \\
    &\qquad + (1 - \ddelay{131})(r^c_{12} + \ddelay{12} r^c_{21}) \Big]
    \\
    &\quad + a_{12} (r^c_{13} + \ddelay{13} r^c_{31}) - a_{13} (r^c_{12} + \ddelay{12} r^c_{21})
    \\
    &\quad + a_{21} [r^c_{13} - (1 - \ddelay{131}) r^c_{12} + \ddelay{13} r_{31}]
    \\
    &\quad - a_{31} [r^c_{12} - (1 - \ddelay{121}) r^c_{13} + \ddelay{12} r^c_{21}]
    \qs
\end{split}
\end{equation}
We see that the former algorithm performs slightly better then the one proposed above, by about a factor of two in amplitude.

This difference in the performance is explained by the factorization used to build the second-generation correction from the first-generation version. The factorized delays are applied to the whole first generation expression, which partly accounts for the the time-varying beatnote frequencies, leading to a lower residual. This is only valid under the assumption that Doppler-delay operators can be commuted. In our setup, this assumption is only an approximation, but the residual due to this effect (i.e., terms scaled by the commutator of two delay operators) is smaller than that due to time-varying beatnotes (i.e., arising from the commutator of beatnote frequencies and delay operators).

A first solution to achieve the same performance in the framework of this paper is to build the clock-noise correction for the first-generation Michelson $X_1$ combination under the same assumptions, and then apply the same factor $(1 - \ddelay{12131})$. Our simulations show that we exactly recover the same levels of residual clock noise. Note, however, that this solution only applies to \gls{tdi} combinations which allow for such a factorization of the second-generation version.

A second, more general solution is to observe that beatnote frequencies always appear directly to the left of the clock noise terms in \cref{eq:q-in-tdi-rearranged}. Going to \cref{eq:q-in-tdi-commuted}, we commuted the $\poly{ij}$ with all beatnote frequencies. A more conservative approach would be to construct expressions $R_{ij}^{a_{ij}}$ which evaluate to
\begin{equation}
    R_{ij}^{a_{ij}} = \poly{ij} a_{ij} q_i
    \qc
\end{equation}
and similarly for the other terms in \cref{eq:q-in-tdi-rearranged}, such that they can be used directly in that equation. This can be approximated by directly rescaling all $r_{ij}$ by $a_{ij}$ before constructing $R_{ij}$. Inspecting our correction variable in \cref{eq:q-in-r}, one of the clock-noise terms always appears attached to a delay operator. This leaves us with a single delay-beatnote commutator in each $r_{ij}$ term, due to the physical propagation delay. Note that this alternative approach can be applied to any \gls{tdi} combination.

The results of this alternative scheme for $X_2$ are presented in \cref{fig:alternative} as the green dashed line. We don't recover the exact zeros from~\cite{Tinto:2018} as they are related to the factor $(1 - \ddelay{12131})$, but otherwise achieve a very similar residual noise level.

\section{Conclusion}
\label{sec:conclusion}

In this article, we revisit the problem of clock-noise correction in space-based \gls{gw} detectors. We find a general expression for the residual clock noise in any \gls{tdi} variable, and provide a generic algorithm to reduce clock noise applicable to most \gls{tdi} variables used in the literature.

We provide expressions for the residual clock noise levels after the correction, and show they are comparable to those previously proposed in the literature~\cite{Tinto:2018}. We study the effect time-varying beatnote frequencies have on these residuals and provide estimates of their magnitude. This effect was neglected in previous studies, and is limiting the residual clock noise with our simulation parameters.

These analytical results are backed up by time-domain simulations using the simulator LISA Instrument. In our studies, the spacecraft positions are simulated using numerical orbits provided by ESA, and the propagation delays are computed including relativistic corrections. In the analytical expressions, we assume that the beatnote frequencies, which are impacted by orbital mechanics, can be described as linear functions of time. This assumption yields excellent agreement with our simulation results.

In order to evaluate clock-noise suppression in comparison to other noise sources, we include a standard \SI{1}{\pico\meter} noise allocation curve. We show that, using realistic values for the on-board \gls{uso}, clock noise is dominating the overall noise level at low to mid frequencies if no clock-noise correction is performed. Using the algorithm we propose, the residual clock noise is suppressed significantly below the required level.

The clock-noise correction algorithm introduces a new modulation noise due to imperfections of the clock sidebands (used for the correction) with respect to the the pilot tone (used as timing reference). We include these noise terms, as well as an algorithm to remove half of them using sideband beatnotes in the reference interferometers. We show that the remaining modulation noise is limiting the achievable noise suppression levels in the clock-corrected observables. Therefore, the signal chain from pilot tone to modulation sideband must fulfill strict timing requirements.

Future work on this topic could include a more realistic model for the clock behaviour, in particular accounting for large deterministic frequency offsets of the onboard oscillators.

\Gls{tdi} specific effects, such as flexing-filtering coupling, coupling of errors in the absolute ranging or in the interpolation scheme, could also be included in our model and simulations. However, their coupling to clock noise is expected to be much less relevant then their impact on the laser noise suppression, due to the much stricter suppression requirements for laser noise. Therefore, we do not expect them to significantly change the results presented here.

\appendix

\section{Geometric TDI algorithm}
\label{sec:geometric-tdi-algorithm}

We give here the algorithm presented in~\cite{Vallisneri:2005}, updated to our notation and to include Doppler shifts. The starting point is a set of one-way measurements $\eta_{ij} = \ddelay{ij} p_j - p_i$. To apply the algorithm, we also need the delay operators $\ddelay{ij}$ and their inverse advancement operators $\ddelay{-ij}$, which fulfill
\begin{equation}
    \ddelay{ij} \ddelay{-ji} x(t) = \ddelay{-ji} \ddelay{ij} x(t) = x(t)
    \qc
\end{equation}
for any function $x(t)$.

The goal is to construct a \gls{tdi} combination which evaluates to
\begin{equation}
    \text{TDI} = \ddelay{\pm i_1j_1} \dots \ddelay{\pm i_n j_n}p_{j_n} - p_{i_1}
    \qc
    \label{eq:geometric-tdi-goal}
\end{equation}
where positive tuples of spacecraft indices ($+ i_k j_k$) represent a delay while negative tuples ($- i_k j_k$) represent an advancement. In order to represent a valid interferometer, adjacent indices must fulfill $j_k = i_{k+1}$. 

The algorithm can be summarized as follows:
\begin{enumerate}
	\item Create a list of the desired time shifts as they appear in \cref{eq:geometric-tdi-goal},
	\begin{equation}
		[\pm i_1 j_i, \dots, \pm i_n j_n]
		\qs
	\end{equation}
	\item Create an empty list $\mathcal{T}$ of delays/advancements, and initialize our \gls{tdi} combination as $\text{TDI} = 0$.
	\item Iterate through the list of time shifts, doing the following operations, assuming the indices are $ij$: if the entry is negative (i.e., an advancement), subtract 
	\begin{equation}
		\text{TDI} \leftarrow \text{TDI} - \mathcal{T} \ddelay{-ij} \eta_{ji}
		\qs
	\end{equation}
	If the entry is positive (i.e., a delay), instead add 
	\begin{equation}
		\text{TDI} \leftarrow \text{TDI} + \mathcal{T} \eta_{ij}
		\qs
	\end{equation}
	Then, append either $\ddelay{-ij}$ or $\ddelay{ij}$ to $\mathcal{T}$.
	\item Now, $\text{TDI}$ evaluates to \cref{eq:geometric-tdi-goal}.
\end{enumerate}


\appendix
\begin{acknowledgments}
We are grateful to the members of the LISA Data Processing Group for their comments and help in improving the manuscript. In particular, we thank M.~Muratore, G.~Heinzel, A.~Petiteau and M.~Tinto for the fruitful discussions regarding this topic. We thank the referees for their help in improving the clarity of this manuscript. We also would like to thank M.~Staab for reviewing the numerical implementation of this algorithm.

We gratefully acknowledge support by the Deutsches Zentrum für Luft- und Raumfahrt (DLR, German Space Agency) with funding from the Federal Ministry for Economic Affairs and Energy based on a resolution of the German Bundestag (Project Ref.~No.~50OQ1601 and 50OQ1801). This research has been supported by Centre National d'Études Spatiales (CNES), Centre National pour la Recherche Scientifique (CNRS), Université Paris-Diderot, as well as the NASA Postdoctoral Fellowship program. The development of LISANode, LISA Instrument and PyTDI is part of the LISA Data Processing Group activities.
\end{acknowledgments}

\bibliographystyle{apsrev4-1}
\bibliography{references}

\end{document}